\documentclass[aps,showpacs,nofootinbib,reprint,pra]{revtex4-1}

\usepackage{latexsym}
\usepackage{graphicx}
\usepackage{graphics}
\usepackage[T1]{fontenc}
\usepackage{multirow}

\newcommand{\ket}[1]{\mbox{$ | #1 \rangle $}}

\newcommand{\equationname}[1]{\textsc{Eq.}~#1}

\begin{document}

\author{A. Martin$^{1}$}
\author{F. Kaiser$^{1}$}
\author{A. Vernier$^{1}$}\email{Currrently with the Laboratoire Charles Fabry, Institut d'Optique Graduate School, Palaiseau, France.}
\author{A. Beveratos$^{2}$}
\author{V. Scarani$^{3,4}$}
\author{S. Tanzilli$^{1}$}\email{sebastien.tanzilli@unice.fr}

\affiliation{
$^1$Laboratoire de Physique de la Matière Condensée, CNRS UMR 7336, Université de Nice -- Sophia Antipolis, Parc Valrose, 06108 Nice Cedex 2, France\\
$^2$Laboratoire de Photonique et Nanostructures, LPN-CNRS UPR20, Route de Nozay, F-91460 Marcoussis, France\\
$^3$Centre for Quantum Technologies, National University of Singapore, 3 Science Drive 2, Singapore 117543\\
$^4$Department of Physics, National University of Singapore, 2 Science Drive 3, Singapore 117542
}

\title{Cross time-bin photonic entanglement for quantum key distribution}

\begin{abstract}
We report a fully fibered source emitting cross time-bin entangled photons at 1540\,nm from type-II spontaneous parametric down conversion.
Compared to standard time-bin entanglement realizations, the preparation interferometer requires no phase stabilization, simplifying its implementation in quantum key distribution experiments.
Franson/Bell-type tests of such a cross time-bin state are performed and lead to two-photon interference raw visibilities greater than 95\%, which are only limited by the dark-counts in the detectors and imperfections in the analysis system. Just by trusting the randomness of the beam-splitters, the correlations generated by the source can be proved of non-classical origin even in a passive implementation. The obtained results confirm the suitability of this source for time-bin based quantum key distribution.
\end{abstract}


\maketitle

\textit{Introduction. --} Entanglement stands as an essential resource for quantum key distribution (QKD)~\cite{gisin_QKD_2002,Scarani_QKD_2009} and quantum communication protocols~\cite{Tittel_photonic_2001}, such as teleportation~\cite{Landry_QtelePlainPalais_2007}, entanglement swapping~\cite{Takesue_Swapp_2009}, relays~\cite{Aboussouan_dipps_2010}, and repeaters~\cite{Sangouard_DLCZRMP_2011}. Long-distance distribution of entangled photons has been demonstrated using various observables, namely polarization~\cite{Aspelmeyer_Polar600m_2003,Ursin_Polar144km_2007}, energy-time~\cite{tittel_experimental_1998}, as well as time-bin~\cite{brendel_pulsed_1999}. Fully operational ``out of the lab'' quantum communication imposes stringent constraints on the source in terms of compactness, stability, and brightness, as well as quality of entanglement. Promising results were obtained using entangled photon sources built around nonlinear integrated waveguide generators~\cite{Tanzilli_genesis_2012}. In this framework, operating wavelengths around 1550\,nm (telecom C-band) offer the possibility of long-distance entanglement distribution, taking advantage of low loss standard optical fibers, as well as high-performance telecommunications components~\cite{Thew_nonmaxTB_2002,Marcikic_TB50km_2004,halder_photon-bunching_2005,takesue_long-distance_2010,Kaiser_typeII_2012}. 

In this paper, we demonstrate a novel time-bin entangled photon-pair source encompassing all the necessary criteria for real-world quantum applications. Unlike previous time-bin source realizations~\cite{tittel_experimental_1998,tanzilli_ppln_2002,Marcikic_TB50km_2004}, our approach exploits cross-polarized paired photons generated from a type-II periodically poled lithium niobate waveguide (PPLN/W)~\cite{Kaiser_typeII_2012}. A polarization to time-bin observable transcriber then enables preparing the photons in the time-bin Bell state $\ket{\Psi^-}_{A,B} = \frac{1}{\sqrt{2}} \left(\ket{0_A,1_B} - \ket{1_A,0_B} \right)$, where $0$ and $1$ denote short and long time-bins, and $A$ and $B$ the parties Alice and Bob, respectively. To our knowledge, the analysis of such a state has never been reported. This scheme, implemented in the continuous wave (CW) regime, is completely insensitive to phase fluctuations in the time-bin preparation stage (the transcriber), and allows 75\% of the detected pairs to be exploited, as opposed to 50\% for standard pulsed regime time-bin schemes~\cite{tittel_experimental_1998,tanzilli_ppln_2002,Marcikic_TB50km_2004}. By simply trusting the randomness of beam-splitters, it provides correlations that cannot be achieved with classical means even in a passive implementation.

\begin{figure}[b]
\centerline{\includegraphics[width=1\columnwidth]{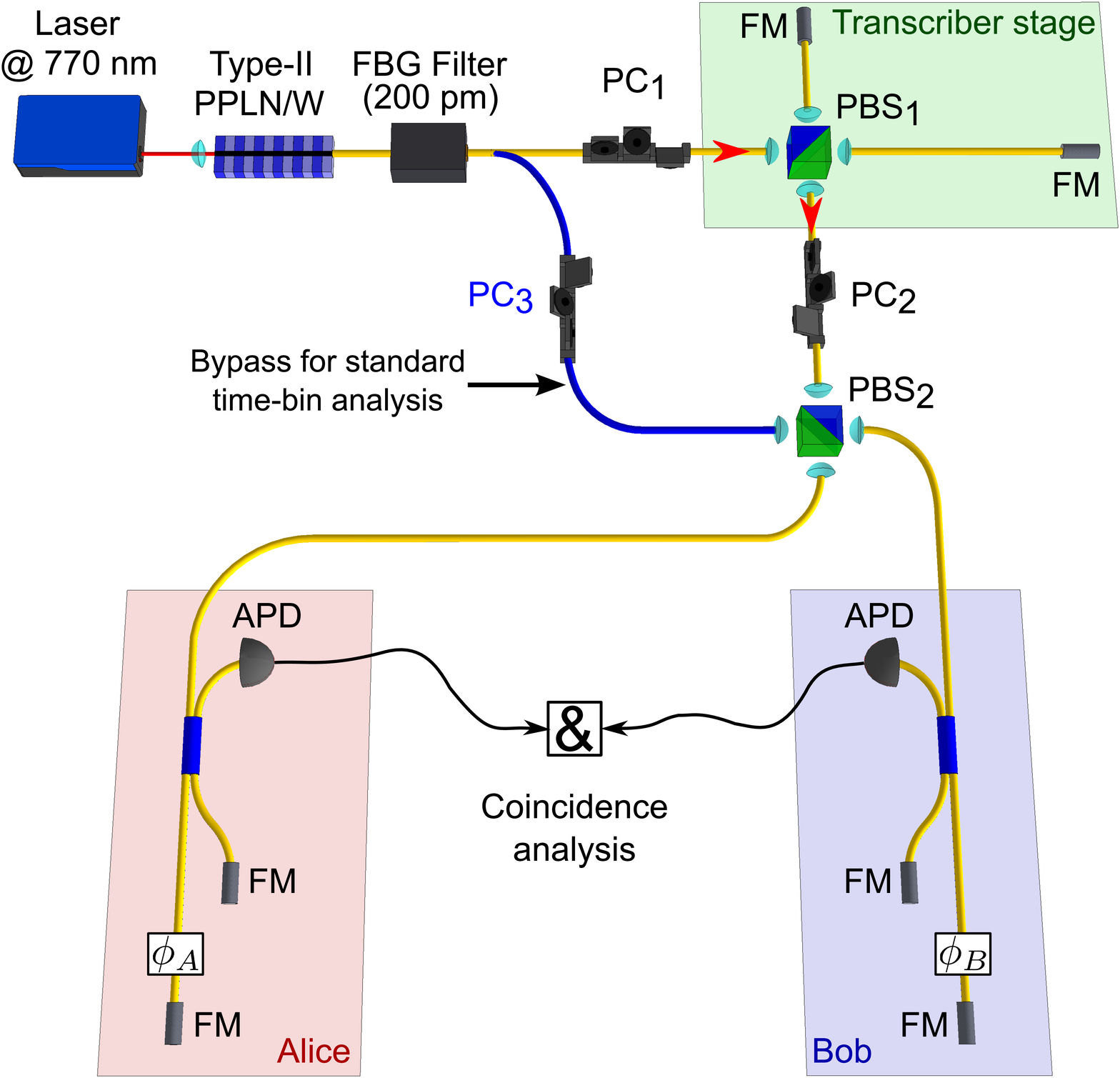}}
\caption{\label{fig_setup}Experimental setup for generating and analysing time-bin entangled states starting with cross-polarized paired-photons. Two different Bell states can be prepared: on one hand the $\ket{\Phi^+}$ state, using the bypass, and on the other hand the $\ket{\Psi^-}$ state, using the transcriber. The analysis is done using two equally unbalanced Mach-Zehnder interferometers in the Franson configuration~\cite{franson_bell_1989}. APD: avalanche photodiode; FM: Faraday mirror.}
\end{figure}

\textit{The setup. --} The setup of the source is shown in \figurename{~\ref{fig_setup}}. Pairs of cross-polarized, \textit{i.e.} horizontally ($|H\rangle$) and vertically ($|V\rangle$), photons are generated at the wavelength of 1540\,nm from spontaneous parametric down-conversion (SPDC) of a 770\,nm CW, $|H\rangle$ polarized, 2.5\,mW, pump laser in a type-II PPLN waveguide. This degenerate phase-matching condition is reached in our case for a 3.6\,cm long, 9.0\,$\mu$m periodically poled, sample, heated at the temperature of 110$^{\circ}$C~\cite{Kaiser_typeII_2012}. Note that various phase-matching conditions obtained in periodically poled materials can be found summarized in~\cite{Tanzilli_genesis_2012}.
The brightness of the PPLN/W was measured to be $\sim 2 \cdot 10^4$ generated pairs per second, mW of pump power, and GHz of emission bandwidth, coupled into a single mode fiber~\cite{Kaiser_typeII_2012}, while multiple-pair emission probability is kept below 1\% per time detection window.
We then select only pairs of wavelength degenerate photons so as to prevent any polarization discernibility as a function of the wavelength~\cite{Kaiser_typeII_2012}. To this end, a fiber Bragg grating (FBG) filter is used to reduce the natural SPDC emission bandwidth from $\sim$850\,pm down to 200\,pm, therefore avoiding undesirable spectral responses associated with alternative phase-matchings. This results in a single-photon coherence time of $\sim$17\,ps ($\leftrightarrow$ coherence length of $\sim$5\,mm).

After the filtering stage, the two photons are sent to a transcriber, arranged as a Michelson interferometer, so as to introduce a time delay between the $\ket{H}$ and $\ket{V}$ polarization modes. A fiber polarization controller (PC$_1$) allows optimizing the separation of the two polarization components at a fiber-pigtailed polarizing beam-splitter (PBS$_1$). In each arm, we use a fiber Faraday mirror (FM), rotating the associated polarization state by 90$^{\circ}$ after a round trip. This ensures that both photons leave the transcriber through the output port of PBS$_1$. For this experiment, the transcriber's optical path length difference is set to $\sim$60\,cm ($\leftrightarrow\,\sim$2\,ns), which is much greater than both the coherence time of the single-photons and timing jitter of the employed detectors ($\sim$0.4\,ns). This prevents the photons to overlap temporally and the state at the output of the transcriber therefore reads $\ket{H,0} \ket{V,1}$, where $0$ and $1$ denote short and long time-bins, respectively.
Next, the photons are sent to an additional polarizing beam-splitter (PBS$_2$) which is oriented at 45$^{\circ}$ with respect to the $\{H;V\}$ basis in which the photons are created.
This way, the polarization modes are no longer associated with the time-bins, therefore reducing the two-photon state to $\ket{\psi}_{A,B} = \frac{1}{2} \left( \ket 0_A \ket 1_A + \ket 0_A \ket 1_B - \ket 1_A \ket 0_B - \ket 1_B \ket 0_B \right)$. Subsequently, the maximally entangled Bell state $\ket{\Psi^-}_{A,B} = \frac{1}{\sqrt{2}} \left( \ket 0_A \ket 1_B - \ket 1_A \ket 0_B \right)$ can be post-selected among all other events using a coincidence detection electronics between the two parties. This state is insensitive to the phase accumulated by the two photons in the transcriber, unlike in standard time-bin schemes where entanglement is prepared using a stabilized unbalanced interferometer placed on the path of a pulsed pump laser~\cite{tanzilli_ppln_2002,Thew_nonmaxTB_2002,Marcikic_TB50km_2004}.

The entanglement analysis is performed using a Franson setup with two equally unbalanced interferometers, one at each side~\cite{franson_bell_1989}. Each interferometer is made of a 50/50 fiber coupler and two fiber-pigtailed FMs to compensate for the birefringence within the analyzer~\cite{tittel_experimental_1998,tanzilli_ppln_2002,Thew_nonmaxTB_2002,Marcikic_TB50km_2004}. As the interferometers' path length difference are adjusted to suitably match the transcriber time-bin separation, such analyzers transform their respective incoming single-photon state in the following way, $\ket \tau_j \mapsto \frac{1}{\sqrt{2}} \left( \ket \tau_j + e^{i \phi_{j}} \ket{\tau+1}_j \right)$, where $\tau = \{0,1\}$ represents the considered time-bin, $j$ Alice or Bob, and $\phi_j$ the phase in Alice's or Bob's interferometer, respectively. After the complete analysis apparatus including the post-selection procedure, the two-photon state reads:
\begin{eqnarray}
\ket \psi&=& \frac{1}{\sqrt{8}} \Big( \ket 0_A \ket 1_B + e^{i \, \phi_A} \ket 1_A \ket 1_B \nonumber\\
&\,&+ e^{i \, \phi_B} \ket 0_A \ket 2_B + e^{i \, (\phi_A + \phi_B)} \ket 1_A \ket 2_B \nonumber\\
&\,&-\ket 1_A \ket 0_B - e^{i \, \phi_A} \ket 2_A \ket 0_B \nonumber\\
&\,&- e^{i \, \phi_B} \ket 1_A \ket 1_B - e^{i \, (\phi_A + \phi_B)} \ket 2_A \ket 1_B \Big) \label{outputstate}.
\end{eqnarray}

\begin{figure}[t!]
\centerline{\includegraphics[width=1\columnwidth]{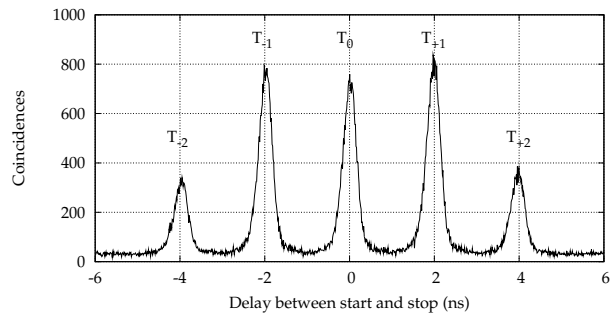}}
\caption{\label{fig_histo}Coincidence histogram for the $\ket{\Psi^-}$ state showing the five expected peaks (see text for details).}
\end{figure}

As can be understood from \equationname{\ref{outputstate}}, one needs to consider five relative arrival times between Alice and Bob detectors. The time-dependent second order intensity correlation function is measured using a free running InGaAs avalanche photodiode (APD, IDQ-210) on Alice's side which is used as a trigger for Bob's gated InGaAs APD (IDQ-201). \figurename{~\ref{fig_histo}} shows the coincidence histogram as a function of the time delay between the photon detection on Alice's and Bob's sides. The central peak, labelled as $\rm T_0$, contains the two contributions associated with photon-pairs in the state $\ket 1_A \ket 1_B$.
The indistinguishablitiy of these two contributions results in usual time-bin entanglement for which the coincidence rate follows $\rm R_c^{T_0} \propto \cos^2 \frac{\phi_A-\phi_B}{2}$. In our CW experimental setup, the pump coherence time ($\tau_c^p \simeq 400\,\rm ns$) is significantly larger than the time-bin separation (2\,ns), such that the emission times of the paired photons remain unknown within $\tau_c^p$. As a result, the paths $\ket 0_A \ket 1_B$ and $\ket 1_A \ket 2_B$, as well as $\ket 1_A \ket 0_B$ and $\ket 2_A \ket 1_B$, are also indistinguishable. These contributions are respectively labelled as $\rm T_{-1}$ and $\rm T_{+1}$ in \figurename{~\ref{fig_histo}}. In these cases, the coincidence rate is given by $\rm R_c^{T_{\pm1}} \propto \cos^2 \frac{\phi_A+\phi_B}{2}$. Remarkably, the peaks $\rm T_{\pm 1}$ on one hand, and $\rm T_{0}$ on the other hand, naturally define two complementary basis useful for entanglement-based QKD protocols~\cite{gisin_QKD_2002,Scarani_QKD_2009,BCHSH_1969,Ekert_QKD_1991}.

 \begin{figure*}[t]
  \begin{tabular}{c c}
    \begin{minipage}{\columnwidth}
      \centering (a) \vspace{-0.2cm} \\
     \includegraphics[width=1\columnwidth]{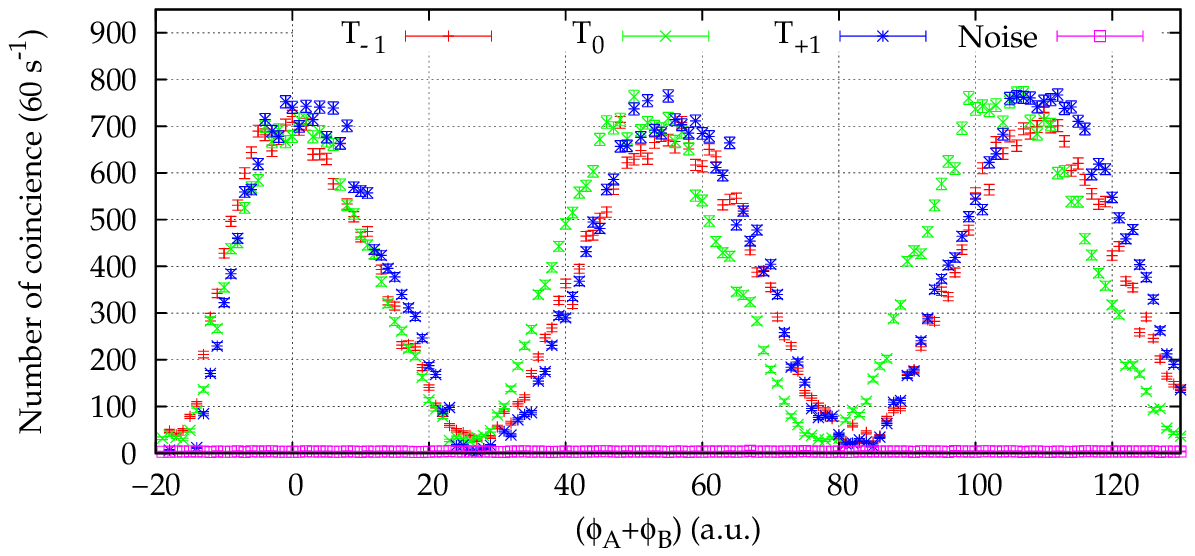}\\
    \end{minipage}&
   \begin{minipage}{\columnwidth}
      \centering (b) \vspace{-0.2cm}\\
     \includegraphics[width=1\columnwidth]{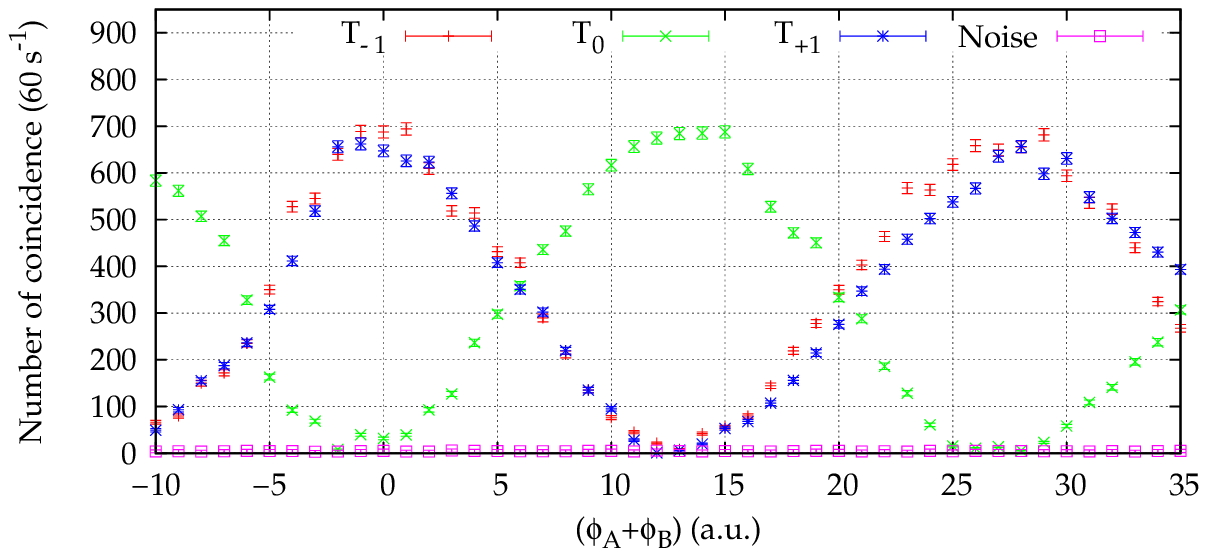}\\
    \end{minipage}
    \vspace{-0.2cm}
  \end{tabular}
 \caption{\label{fig_bell_psi_2}Bell tests performed for the $\rm T_{\pm1}$ and $\rm T_{0}$ conincidence peaks of \figurename{~\ref{fig_histo}} as a function of $\phi_A + \phi_B$, with $\phi_A$ continuously tuned and $\phi_B$ set to 0 (a) and to $\pi/2$ (b).}
  \vspace{-0.5cm}
 \end{figure*}

To quantify the quality of our analysis interferometers, we first perform a two-photon interference experiment using a genuine energy-time Franson setup~\cite{franson_bell_1989}. In this case, the transcriber is bypassed (see \figurename{~\ref{fig_setup}}), such that the cross-polarized pairs of photons are directly sent to PBS$_2$. The polarization state of the two photons is adjusted using PC$_3$ for the pairs to be deterministically separated at this PBS. The relative phase between the analysers is varied by tuning the temperature of Alice's interferometer, while that of Bob's is kept constant. We achieve net and raw visibilities of 98$\pm$2\% and 96$\pm$2\%, respectively (curve not represented).
The net visibility is obtained after subtraction of accidental coincidences due to the detectors dark-counts.
We ascribe $\sim$1\% net contrast reduction to the length mismatch between the two analysers which was measured to be 0.3$\pm$0.1\,mm. Note that multiple-pair generation probabilities in this experiment are very small ($<$0.1\%).

Eventually, we test the cross time-bin state $\ket{\Psi^-}_{A,B}$ prepared with the transcriber. For the three coincidence peaks $\rm T_{-1}$, $\rm T_{+1}$, and $\rm T_{0}$, two-photon interference is also recorded by tuning the phase $\phi_A$, while keeping $\phi_B \approx 0$. As shown in \figurename{~\ref{fig_bell_psi_2}} (a), the three coincidence rates are modulated as a function of the global phase ($\phi_A + \phi_B$) and lead to net (raw) visibilities exceeding 97$\pm$2\% (95$\pm$2\%). For this test, we ascribe an additional 1\% visibility reduction to the path length mismatch between the transcriber and the analysers. In addition, the visibility associated with events detected in the $\rm T_0$ time-bin is strongly dependent on the polarization state adjustments in front of the second PBS where the $\ket{\Psi^-}$ state is prepared. Also, the raw visibilities could be improved by employing detectors showing better dark-count figures, such as those offered by super-conducting devices. 
Furthermore, to show that the three contributions follow the expected phase dependency, we change the phase of Bob's interferometer to $\phi_B = \pi/2$. As expected, \figurename{~\ref{fig_bell_psi_2}} (b) confirms that the phase relation of the interference patterns between $\rm T_{\pm1}$ and $\rm T_{0}$ is shifted by $\pi$.
Note that the analysers are only temperature stabilized which explains the drifts between the patterns shown in \figurename{~\ref{fig_bell_psi_2}} (a) and (b). This could be avoided by employing active interferometer stabilization schemes enabling accurate phase control and setting, as was demonstrated using a frequency stabilized laser combined to a phase measurement~\cite{DeRiedmatten_LDswapp_2005,Kaiser_Type0_2013}.

\textit{Theory. --} For the characterization of the setup, we have plotted interference patterns, out of which one may check the violation of some Bell inequality~\cite{BCHSH_1969} under some assumptions proper to time-bin implementations~\cite{Aerts_FransonLocalRealism_1999,Cabello_GenuineET_2009}. However, we want to operate this setup for QKD in a \textit{passive implementation}, that is $\phi_A$ and $\phi_B$ are going to be fixed. The detections of any passive setup can always be reproduced by a local variable model if one does not make additional assumptions: indeed here, if Eve can choose the time of the detection, she can obviously fulfill all the requirements. So we make the following assumption on Alice's and Bob's measurements separately: if some signal comes in the analyzer at time $\tau$, the detection can happen either at $\tau$ or at $\tau+1$ (up to the propagation time in the interferometer) and this time is \textit{not} for Eve to control. In other words, we trust the randomness introduced by the beam-splitter. Under this assumption, we are going to provide a semi-device-independent criterion for quantumness, analog to a Bell inequality~\cite{note1}.

Consider Eve sending some signal in Alice's lab at time $\tau_A$ and some signal in Bob's lab at times $\tau_B$. As per our assumption, detection at Alice's side can happen at $\tau_A$ or at $\tau_A+1$. Apart from this uncertainty in the time of firing, we assume the most favorable situation for Eve, namely that for both times she knows with certainty which detector will fire; we also assume that only one detector will fire (notice that there may be no state of light such that this can actually happen: we give more power to Eve by assuming \textit{only} her impossibility of choosing the time of the firing). We do similar assumptions for Bob. Thus, Eve knows the four possible outcomes $a(\tau_A)$, $a(\tau_A+1)$, $b(\tau_B)$, $b(\tau_B+1)$, with $(a,b)=\pm1$.

For definiteness, we assume $\phi_A=\phi_B=0$: for these values, ideally, the quantum source produces identical outcomes at $\rm T_{0}$ and $\rm T_{\pm1}$, and completely uncorrelated outcomes at $\rm T_{\pm2}$. The QKD protocol we want to propose uses the detections at $\rm T_{0}$ and $\rm T_{\pm1}$ as the raw key, while the lateral peaks could be used to detect the presence of an eavesdropper. We need to see how close Eve can come to simulating the ideal quantum source (we assume that the beam-splitters in the Michelson interferometer have 50/50 coupling ratio, although this additional assumption can easily be relaxed):
\begin{itemize}
\item[(a)] $\tau_A=\tau_B=\tau$: in this case, $\rm T_{0}$ happens with probability $\frac{1}{2}$, and $\rm T_{\pm1}$ with probability $\frac{1}{4}$ each. Eve can ensure that Alice and Bob get the same outcomes by fixing $a(\tau)=a(\tau+1)=b(\tau)=b(\tau+1)$. But, if Eve does only this, there will never be any event $\rm T_{\pm2}$.
\item[(b)] $\tau_A=\tau_B-1=\tau$: in this case, $\rm T_{+1}$ happens with probability $\frac{1}{2}$, and $\rm T_{0,+2}$ with probability $\frac{1}{4}$ for each. Now, Eve cannot guarantee all the conditions: Eve can set $a(\tau)=b(\tau+1)=a(\tau+1)$ without loss of generality, but she has to choose whether $b(\tau+2)$ is also equal to those, so as to get the case $(\tau+1,\tau+2)$ right; or random, to get the case $(\tau,\tau+2)$ right.
\item[(c)] $\tau_A=\tau_B-2=\tau$: in this case, $\rm T_{+2}$ happens with probability $\frac{1}{2}$, and $\rm T_{+1,+3}$ with probability $\frac{1}{4}$ for each. Eve can set $a(\tau+1)=b(\tau+2)$, while $a(\tau)$ and $b(\tau+3)$ are uncorrelated with anything. This way, she gets all the expected detections, but she may generate an unexpected event at $\rm T_{3}$.
\end{itemize}
As we see, one has to check for high correlations in $\rm T_{0,\pm1}$, for the presence of uncorrelated coincidences in $\rm T_{\pm2}$, and for the absence of coincidences in $\rm T_{\pm3}$. There are several ways to do that; among the simplest, one can form a single linear function like
\begin{eqnarray}
S&=& \sum_{x = 0,\pm 1}P({\rm T_x})E_{{\rm T_x}}\,+\,2\,\sum_{x = \pm 2}P({\rm T_x})(1-E_{{\rm T_x}})\nonumber\\
&&-\,\sum_{x = \pm 3}P({\rm T_x})
\end{eqnarray} where $P({\rm T_x})$ is the probability of a coincidence $\rm T_x$ and $E$ is the correlation coefficient. All the classical strategies above reach at most $S=1$, while perfect experimental results would lead to $S=1.25$. Our experimental raw data from \figurename{~\ref{fig_bell_psi_2}} (a) lead to $S_{raw}^{exp}=1.20\pm0.02$, corresponding to 10 standard deviations above the classical boundary.
To infer this value, we take into account the above mentioned visibilities for the coincidence peaks $\rm T_{\pm1}$ and $\rm T_{0}$, as well as an average visibility of 0$\pm$2\% for the coincidence peaks $\rm T_{\pm2}$ (curves not shown).
This result can be improved by employing lower dark-count detectors and by adjusting actively the path length difference of the analysis interferometers to match that of the transcriber. Note that a similar theoretical/experimental analysis with $\phi_A=-\frac{\pi}{2}$ and $\phi_B=\frac{\pi}{2}$ (case of \figurename{~\ref{fig_bell_psi_2}} (b)) leads to the same results and conclusion. In this case, Alice and Bob expect the observation of coincidences in $\rm T_0$ and simultaneously anti-coincidences in $\rm T_{\pm 1}$. As Eve cannot know in which coincidence peak the photons are detected, she cannot mimic the correct coincidence/anti-coincidence behavior.

\textit{Conclusion. --} We have demonstrated, for the first time, a scheme that allows analysing the cross time-bin entangled state $\ket{\Psi^-}$. From the experimental side, the simplicity of the scheme associated with the phase-insensitivity of the prepared state to the transcriber fluctuations, make this source a promising candidate for time-bin entanglement based long-distance QKD. In this perspective, such a novel strategy offers naturally the two necessary complementary basis. This makes it possible to operate entanglement based QKD protocols~\cite{Ekert_QKD_1991,Ling_QKDBell_2008} in a full passive fashion, requiring only one stabilized analyser plus two detectors at each user's location, and allowing the exploitation of 75\% of the detected pairs of photons~\cite{Tittel_TimBinQKD_2000}.
Moreover, the source can be operated in both CW and pulsed regimes, without any further experimental modification but the laser. The latter regime is particularly interesting when synchronization between different locations is required, as is the case in quantum relay schemes~\cite{Aboussouan_dipps_2010}. Using exclusively standard fiber telecom components allowed us reaching an overall loss figure of $\sim$5\,dB from the output of the PPLN/W to the detectors.
A simple theoretical model trusting only the randomness of the beam-splitters also confirms the quantum nature of the source, and therefore its potential for passive implementations of entanglement-based QKD.

We thank W. Sohler and H. Herrmann for providing the PPLN waveguide and for fruitful discussions. Financial support is acknowledged from the CNRS, the Conseil Régional PACA, the French Minist\`ere de l'Enseignement Sup\'erieur et de la Recherche (MESR), the European program ERA-SPOT ``WASPS'', the iXCore Research Foundation, as well as from the Ministry of Education and the National Research Fund of Singapore.

\textit{Note added: during the preparation of this manuscript, we became aware of a preprint paper in which cross time-bin states are employed for studying, in a convincing manner, 'time-bin entangled photon holes'~\cite{Liang_QHolesTB_2012}.}


%

\end{document}